\documentclass[twocolumn]{aastex631}

\usepackage{xcolor}
\usepackage{apjfonts}
\usepackage{graphicx}
\usepackage{epstopdf}
\usepackage{ulem}
\usepackage[utf8]{inputenc}
\usepackage[T1]{fontenc}
\usepackage{url}
\usepackage{mdwlist}
\usepackage{amsmath} 
\usepackage{enumitem}

\begin{document}

\newcommand{\ls}{{_<\atop^{\sim}}}
\newcommand{\gs}{{_>\atop^{\sim}}}
\def \spose#1{\hbox  to 0pt{#1\hss}}  
\def \ls{\mathrel{\spose{\lower 3pt\hbox{$\sim$}}\raise  2.0pt\hbox{$<$}}}
\def \gs{\mathrel{\spose{\lower  3pt\hbox{$\sim$}}\raise 2.0pt\hbox{$>$}}}
\newcommand{\Ha}{\hbox{{\rm H}$\alpha$}}
\newcommand{\Hb}{\hbox{{\rm H}$\beta$}}
\newcommand{\Ovi}{\hbox{{\rm O}\kern 0.1em{\sc vi}}}
\newcommand{\OIII}{\hbox{[{\rm O}\kern 0.1em{\sc iii}]}}
\newcommand{\OII}{\hbox{[{\rm O}\kern 0.1em{\sc ii}]}}
\newcommand{\lya}{Ly$\alpha$}
\newcommand{\NII}{\hbox{[{\rm N}\kern 0.1em{\sc ii}]}}
\newcommand{\SII}{\hbox{[{\rm S}\kern 0.1em{\sc ii}]}}
\newcommand{\angstrom}{\textup{\AA}}
\newcommand\ionn[2]{#1$\;${\scshape{#2}}}
\newcommand{\fesc}{$ f_{\rm esc}$}
\newcommand{\flya}{$ f_{\rm esc}^{Ly\alpha}$}



\title{Early Results from GLASS-JWST XXII: Rest frame UV-optical spectral properties of Lyman-alpha emitting galaxies at 3 $<$ z $<$ 6}

\shorttitle{Spectral properties of Ly$\alpha$ emitting galaxies at z$>$3 }
 
\shortauthors{Roy et al.}


\correspondingauthor{Namrata Roy}
\email{nroy13@jhu.edu}

\author[0000-0002-4430-8846]{Namrata Roy}
\affiliation{Center for Astrophysical Sciences, Department of Physics and Astronomy, Johns Hopkins University, Baltimore, MD, 21218}

\author[0000-0002-6586-4446]{Alaina Henry}
\affiliation{Space Telescope Science Institute, 3700 San Martin Drive, Baltimore MD, 21218} 
\affiliation{Center for Astrophysical Sciences, Department of Physics and Astronomy, Johns Hopkins University, Baltimore, MD, 21218}

\author[0000-0002-8460-0390]{Tommaso Treu}
\affiliation{Department of Physics and Astronomy, University of California, Los Angeles, 430 Portola Plaza, Los Angeles, CA 90095, USA}

\author[0000-0001-5860-3419]{Tucker Jones}
\affiliation{Department of Physics and Astronomy, University of California Davis, 1 Shields Avenue, Davis, CA 95616, USA}

\author[0000-0003-3518-0374]{Gonzalo Prieto-Lyon}
\affiliation{Cosmic Dawn Center (DAWN), Denmark}
\affiliation{Niels Bohr Institute, University of Copenhagen, Jagtvej 128, DK-2200 Copenhagen N, Denmark}

\author[0000-0002-3407-1785]{Charlotte Mason}
\affiliation{Cosmic Dawn Center (DAWN), Denmark}
\affiliation{Niels Bohr Institute, University of Copenhagen, Jagtvej 128, DK-2200 Copenhagen N, Denmark}

\author[0000-0001-6670-6370]{Tim Heckman}
\affiliation{Center for Astrophysical Sciences, Department of Physics and Astronomy, Johns Hopkins University, Baltimore, MD, 21218}

\author[0000-0003-2804-0648 ]{Themiya Nanayakkara}
\affiliation{Centre for Astrophysics and Supercomputing, Swinburne University of Technology, PO Box 218, Hawthorn, VIC 3122, Australia}

\author[0000-0001-8940-6768]{Laura Pentericci}
  \affil{INAF - Osservatorio Astronomico di Roma, via di Frascati 33, 00078 Monte Porzio Catone, Italy}

\author[0000-0002-9572-7813]{Sara Mascia}
\affiliation{INAF - Osservatorio Astronomico di Roma, via di Frascati 33, 00078 Monte Porzio Catone, Italy}

\author[0000-0001-5984-0395]{Maru{\v s}a Brada{\v c}}
\affiliation{University of Ljubljana, Department of Mathematics and Physics, Jadranska ulica 19, SI-1000 Ljubljana, Slovenia}
\affiliation{Department of Physics and Astronomy, University of California Davis, 1 Shields Avenue, Davis, CA 95616, USA}

\author[0000-0002-5057-135X]{Eros Vanzella}
\affiliation{INAF -- OAS, Osservatorio di Astrofisica e Scienza dello Spazio di Bologna, via Gobetti 93/3, I-40129 Bologna, Italy}

\author[0000-0002-9136-8876]{Claudia Scarlata}
\affiliation{School of Physics and Astronomy, University of Minnesota, Minneapolis, MN, 55455, USA}

\author[0000-0003-4109-304X]{Kit Boyett}
\affiliation{School of Physics, University of Melbourne, Parkville 3010, VIC, Australia}
\affiliation{ARC Centre of Excellence for All Sky Astrophysics in 3 Dimensions (ASTRO 3D), Australia}

\author[0000-0001-9391-305X]{Michele Trenti}
\affiliation{School of Physics, University of Melbourne, Parkville 3010, VIC, Australia}
\affiliation{ARC Centre of Excellence for All Sky Astrophysics in 3 Dimensions (ASTRO 3D), Australia}

\author[0000-0002-9373-3865]{Xin Wang}
\affil{School of Astronomy and Space Science, University of Chinese Academy of Sciences (UCAS), Beijing 100049, China}
\affil{National Astronomical Observatories, Chinese Academy of Sciences, Beijing 100101, China}
\affil{Institute for Frontiers in Astronomy and Astrophysics, Beijing Normal University,  Beijing 102206, China}


\begin{abstract}

Ly$\alpha$ emission  is possibly the best indirect diagnostic of Lyman continuum (LyC) escape since the conditions that favor the escape of Ly$\alpha$  photons are often the same that allows for the escape of LyC photons. 
In this work, we present the rest UV-optical spectral characteristics of 11 Ly$\alpha$ emitting galaxies at 3 $<$ z $<$ 6 - the optimal redshift range chosen to avoid the extreme IGM attenuation while simultaneously studying galaxies close enough to the epoch of reionization. From a combined analysis of JWST/NIRSpec and MUSE data,  we present the Ly$\alpha$ escape fraction and study their correlations with other physical properties of galaxies that might facilitate Ly$\alpha$ escape. We find that our galaxies have low masses (80\% of the sample with $\rm log_{10} \ M_{\star} < 9.5\ M_{\odot}$), compact sizes (median $\rm R_e \sim 0.7 \ kpc $), low dust content, moderate [OIII]/[OII] flux ratios (mean $\sim$ 6.8 $\pm$ 1.2), and moderate Ly$\alpha$ escape fraction (mean $\rm f_{esc}^{Ly\alpha} \ \sim$ 0.11). Our sample show characteristics that are broadly consistent with the low redshift galaxies with Ly$\alpha$ emission, which are termed as ``analogs'' of high redshift population. We predict the Lyman continuum escape fraction in our sample to be low (0.03-0.07), although larger samples in the post-reionization epoch are needed to confirm these trends. 

\end{abstract}



\section{Introduction} \label{sec:Introduction}

One of the main science drivers behind the James Webb Space Telescope (JWST) is spectroscopic confirmation and characterization of the galaxies formed during the crucial epoch of Cosmic Reionization (EoR), i.e., when the intergalactic medium (IGM) transforms from neutral to completely ionized state \citep[see][ for a review]{robertson22}.  The Lyman-continuum (LyC,  $\lambda <$ 912 \AA) photons escaping  into the neutral IGM from the star-forming galaxies can reionize the Universe by z = 6 only if a substantial fraction ($\sim$ 10 \%) of the photons escape from the galaxies' interstellar and circumgalactic media \citep[ISM and CGM; e.g., ][]{finkelstein19, robertson15}. Early data from JWST has revealed insights into the ISM conditions and ionizing photon production in reionization-era galaxies at z $>$ 6 \citep{curtislake22, robertson22b, cameron23, trump22, tacchella22, fujimoto23}. 
However, directly estimating the escape fraction of the LyC photons (\fesc) is almost impossible at z $>$ 4.5 since the photons get absorbed by the dense neutral IGM along the line of sight \citep{inoue14}. 


While a number of indirect tracers have been used to predict LyC \fesc -- e.g., [SII] emission deficit, Mg II emission, high [OIII]/[OII] emission \citep{zackrisson13, nakajima14, wang21, katz22, flury22}, one of the best indicators of LyC \fesc \ is the \lya \ emission  -- the brightest nebular recombination line of hydrogen atom \citep[e.g.,][]{pahl21, gazagnes20}.  LyC leakers tend to show strong \lya \  emission escape  \citep{nakajima20}. Since Ly$\alpha$ scatters resonantly in \ion{H}{1}, the resultant emission profile bears information about the neutral hydrogen in the IGM and the epoch of reionization \citep{stark10, treu12}, as well as the gas covering fraction, column density, and dust geometry of the host galaxy from which the intrinsic \lya \ emission emerges \citep{neufeld91}. 
Since \lya \  remains the strongest diagnostic and has now been observed in some of the highest redshift galaxies using JWST \citep{saxena23, bunker23, boyett23, mascia23}, it is now essential to understand the physical processes that determine \lya\ escape, and to disentangle the effects of the ISM and IGM.

Over the last several years, studies have aimed to understand the escape of \lya\ emission from galaxies at redshifts where the IGM is fully ionized. With the union of HST/ Cosmic Origins Spectrograph (COS) and the SDSS, samples of low-redshift ($z \ls\ 0.5)$ analog galaxies have been studied with detailed spectroscopic coverage from the rest-frame UV to optical \citep[blueberries, green peas, Lyman break analogs, Lyman continuum emitters; ][]{cardamone09, heckman11, heckman15, yang17a, yang17b, henry15, jaskot13, hayes23, flury22}. Diagnostics to predict \lya\ properties have begun to emerge: \cite{hayes13, hayes14} show how the spatial extent of the scattered \lya\ emission varies with galaxy properties. Other studies investigate the escape fraction of \lya\ photons and the galaxy characteristics that promote strong \lya\ emission \citep{henry15, rivera-thorsen15, yang17a}.  Moreover, Mg II emission has been shown to correlate strongly with \lya, further indicating low column densities in the ISM \citep{henry18, chisholm20, xu22, xu23}. And, importantly, low-redshift analogs have been used to develop predictors for LyC escape \citep{heckman11, izotov16, izotov18, flury22}, demonstrating the strong link between LyC and \lya.

However, a critical question remains: \textit{are these diagnostics and correlations applicable in the reionization epoch?} The answer to this question bears heavily on our interpretation of the high-redshift galaxy samples now being uncovered with JWST.  An initial spectroscopic census of JWST galaxies above z $>$ 7 seems to indicate elevated ionization parameters, low metallicities, low dust content, high specific star formation rates (sSFR), and higher ionizing radiation escape fraction  than seen locally \citep{katz23, curti23, cameron23, sanders23}.   While many of these galaxies resemble the extreme ISM conditions observed in low-redshift analogs, we cannot be certain if the similarities extend to  \lya\ and LyC.  Therefore, the critical next step is to verify our low-z diagnostics in the post-reionization epoch. JWST’s access to the rest-frame optical spectra of galaxies at z$\sim 3-6$ now provides the complete spectroscopic coverage needed to make this possible. 


In this paper, we focus on \lya.  Since nearly all low-z analog galaxies show strong \lya\ emission, their most obvious counterparts at $z>3$ are \lya-selected galaxies. Therefore, we present 
a rest frame UV-optical spectroscopic study of 11 \lya \ emitting galaxies (LAEs) at 3$<$z $<$ 6, selected using the Multi-Unit Spectroscopic Explorer (MUSE) instrument at the ESO-VLT. The redshift range is optimally chosen to include moderate to high redshift galaxies, not exceeding beyond z $>$ 6.5, where \lya \ emission gets significantly attenuated due to the IGM \citep{pentericci18, fuller20}. We utilize the JWST/NIRSpec spectra obtained in the Abell 2744 cluster field as part of the JWST Early Release Science program GLASS \citep{treu22} to confirm spectroscopic redshifts and measure rest frame optical line ratios and EWs with high precision. We look for the possible evolution of the \lya\ output of galaxies by comparing to low-z analogs galaxies observed with COS. 
 Although our sample of 11 sources is not large, our objects provide a critical benchmark for future comparisons as more and more JWST data become available over the coming years.

In \S 2, we describe the observations used in this study and the chosen sample. In \S 3, we describe the analysis methods and measurements of the rest frame UV-optical spectral properties that we use throughout this work. In \S 4 \& 5, we present the main results. 
In \S\ref{discussions}, we discuss the implications of the result and end with a conclusion in \S\ref{summary}.

\section{Data Acquisition and Sample Definitions } 

This work focuses on 11 \lya \  emitting sources in the field of the lensing cluster Abell 2744. Our study uses spatially resolved optical IFU spectroscopic data from VLT/MUSE and deep near-infrared spectroscopy from JWST/NIRSpec to obtain rest frame UV-optical spectra. The final list of targets in our sample and their properties are listed in Table~\ref{tab:1}, and the details of the sample selection are discussed in the sections to follow. 

\subsection{VLT/MUSE spectroscopy} \label{muse}

Optical spectroscopy using the VLT/MUSE instrument was performed on the Abell 2744 cluster region as part of the GTO program 094.A-0115 \citep[PI: Richard;][]{mahler18, richard21}{}. The program targeted the central regions of the massive cluster using the wide field mode, with a 2 arcmin $\times$ 2 arcmin mosaic of MUSE pointings. The wavelength coverage of the observations ranges between 4750 to 9350 \AA, with a spectral resolution varying between R = 2000 - 4000. The reduced datacubes are publicly available in the form of fits files\footnote{\hyperlink{http://muse-vlt.eu/science/a2744}{http://muse-vlt.eu/science/a2744}}. \cite{mahler18} published the full spectroscopic catalog of objects with measured redshifts. \cite{richard21} published a follow-up catalog of Ly$\alpha$ emission line measurements, with an emission line detection limit of  (0.77-1.5) $\times \rm \  10-18 \  erg  \ s^{-1} \ cm^{-2}$ at 5$\sigma$. This parent catalog of MUSE LAEs was used to select a subset of 17 LAEs that were observed with NIRSpec as part of the GLASS-JWST program.

\subsection{JWST/NIRSpec spectroscopy} \label{jwst}

This work uses near-infrared spectroscopic data in the Abell 2744 field obtained from the GLASS-JWST Early Release Science (ERS) program \citep[ID 1324, PI Treu;][]{treu22}{}.
This observation provides
high-resolution spectroscopy (R = 2700) over an observed wavelength range of $\rm \lambda_{obs} \sim $ 0.8 - 5.2 $\mu$m using the NIRSpec multi-object spectroscopy mode. The GLASS-JWST observations were carried out with three spectral configurations: G140H/F100LP (0.97 - 1.82 $\mu$m), G235H/F170LP (1.66 - 3.05$\mu$m) and G395H/F290LP (2.87 - 5.14 $\mu$m). Each of the three high-resolution gratings was exposed for 4.9 hours. The MSA target selection is discussed in \cite{treu22}. All the  {\it JWST} data used in this paper can be found in MAST: \dataset[10.17909/fqaq-p393]{http://dx.doi.org/10.17909/fqaq-p393}.

We use the official STScI JWST pipeline\footnote{\hyperlink{https://github.com/spacetelescope/jwst}{https://github.com/spacetelescope/jwst}} (version 1.8.2) and the msaexp code\footnote{\hyperlink{https://github.com/gbrammer/msaexp}{https://github.com/gbrammer/msaexp}} with the updated set of reference files that include in-flight flux calibrations (CRDS\_CONTEXT = "jwst\_1041.pmap") to produce Level 2 and 3 products. \cite{morishita22}  have discussed the data reduction process and observation strategy in detail. In short, we downloaded the Level 1b data products from the MAST portal\footnote{\hyperlink{https://mast.stsci.edu/portal/Mashup/Clients/Mast/Portal.html}{https://mast.stsci.edu/portal/Mashup/Clients/Mast/Portal.html}}.  \texttt{calwebb\_detector1}, which is the first step of the reduction, was already run on the raw detector exposures and are provided in the Level 1b outputs. We implement the second and third steps of the reduction: \texttt{calwebb\_spec2} and \texttt{calwebb\_spec3} routines to perform flat-fielding, wavelength calibrations, path-loss corrections, and background subtractions. Nodded observations of our MSA slitlet exposures were used to
 perform local background subtraction since our sources are compact. 
 The resultant 2D spectra are visually inspected, and the 1D spectra are optimally extracted, following the method outlined by \cite{horne86}. 

The high spectral resolution and the broad wavelength coverage of NIRSpec enable detection of the standard rest-frame optical emission lines like [OII]$\lambda \lambda$3727, 3729 \ \AA, H$\beta$, [OIII]$\lambda \lambda$4959, 5007 \ \AA, and H$\alpha$ up to z $\sim $ 7, allowing for precise measurements of optical spectral properties of our sources. 

\subsection{Sample selection} \label{sample}

A sample of 17 MUSE LAEs with confident Ly$\alpha$ redshifts were included in the NIRSpec MSA configuration. This sample is not magnitude complete but provides an ideal sample to compare with the nearby UV-bright green pea population \citep{henry15, yang17a}, extreme emission line galaxies \citep{erb16}, Lyman Break Analogs \citep{heckman15}, Lyman continuum leaking galaxies \citep{izotov16} and starbursts \citep{rivera-thorsen15, rivera-thorsen17} in the literature, which are believed to be low redshift analogs of high-z LAEs. 12 of those 17 LAEs show strong rest frame optical line detection whose redshifts and spectral properties can be measured with confidence \citep{mascia23}, Prieto-Lyon et al. 2023 (in prep). One of these 12 shows very faint Ly$\alpha$ detection with insufficient S/N below 3.  We use the remaining 11 LAEs as the core sample for this work (Table.~\ref{tab:1}). Our sample is identical to Prieto-Lyon et al. 2023 (in prep), who focus on the \lya \ velocity offsets.

\subsection{Catalogs} \label{catalog}

We use the redshifts from \cite{mahler18} catalog to primarily select our 11 Ly$\alpha$ emitting galaxies at z $>$ 3. The catalog is publicly available \footnote{\hyperlink{http://data.muse-vlt.eu/A2744/A2744\_redshifts\_cat\_final.txt}{http://data.muse-vlt.eu/A2744/A2744\_redshifts\_cat\_final.txt}}. We used the UV-rest frame continuum measurements from \cite{richard21} catalog in our analyses.
The UV continuum is derived using the photometric catalogs for Abell 2744, which includes 
photometry information from the HST/ACS 
and WFC3/IR of the Hubble Frontier Fields project \citep[HST-GO/DD-13495; ][]{lotz17}{}.
We use the \texttt{ASTRODEEP} catalog \footnote{\hyperlink{http://www.astrodeep.eu/frontier-fields/}{http://www.astrodeep.eu/frontier-fields/}} from \cite{merlin16}, and \cite{castellano16} for reporting the stellar mass and star formation rates of our targets.   The stellar
238 mass and star formation rate estimates are derived from SED
239 fitting, taking into account the nebular emission line contribution. See \cite{castellano16} for details.

\subsection{Comparison Sample} \label{comparisonsample}

We aim to test whether the low redshift ($z\ls$0.4) analogs of high-z LAEs are similar to the LAEs at a cosmologically significant redshift (z$>$ 3) in terms of rest frame optical spectral properties. 
But first, we place our high z sample in the context of the broader z$\sim$ 3-6 LAE population previously reported in the literature. 
To select an emission line survey for z = 3-6 LAEs to be compared with our sample, 
we  use the publicly available MUSE 
observations taken in the Hubble Ultra Deep Field (HUDF) region \citep{bacon22, inami17}. 
We chose this survey since this is the deepest spectroscopic
survey ever performed,  and the catalog contains the most comprehensive measurements we require to compare with our high redshift sources. The catalog 
is available through the CDS/ Vizier database or the MUSE data products website \footnote{\href{https://amused.univ-lyon1.fr/project/UDF/HUDF/}{https://amused.univ-lyon1.fr/project/UDF/HUDF/}}.

For the low redshift analog population, we use \cite{hayes23} as our primary sample, which compiles the HST/COS observations of  87 UV-bright galaxies. 
The COS data are drawn from: 
GO 11522 (PI: Green), 11727 (Heckman), 12027 (Green), 12269 (Scarlata), 12583 (Hayes), 12928 (Henry), 13017 (Heckman), 13293 (Jaskot), 13744 (Thuan), 14080 (Jaskot), 14201 (Malhotra), 14635 (Izotov), 15136 (Izotov),  15639 (PI: Izotov), and 15865 (PI: Henry). 
This comprehensive sample includes sources that are frequently classified as low-redshift analogs of high-z galaxies  due to their strong \lya \ emission and emission line properties - like the Green pea population, Lyman continuum leaking galaxies, extreme emission line galaxies, Lyman break analogs, and compact starbursts \citep{heckman11, heckman15, wofford13, henry15, rivera-thorsen15, rivera-thorsen17, izotov16, izotov18, jaskot17, jaskot19, yang17a, izotov21, xu22}. The corresponding rest-optical data are obtained from the Sloan Digital Sky Survey (SDSS) Data Release 16 \citep[DR16; ][]{ahumada20}{}.  The \lya \  and optical emission line properties measurements for this extensive sample are obtained from \cite{hayes23} via private communication. These galaxies are in the redshift range between z = 0.020 and z = 0.44 and sample a large range in stellar mass and star formation rates. We refer to this compilation of galaxies as low-z analogs. 
In addition to the sample described above, we include 88 Lyman continuum emitter candidates from the low z Lyman continuum survey (LzLCs) sample \citep{flury22}. 50\% of these galaxies are confirmed LyC leakers. All galaxies in this sample exhibit strong Ly$\alpha$ emission. 

\begin{figure*}
    \centering
    \includegraphics[width=\textwidth]{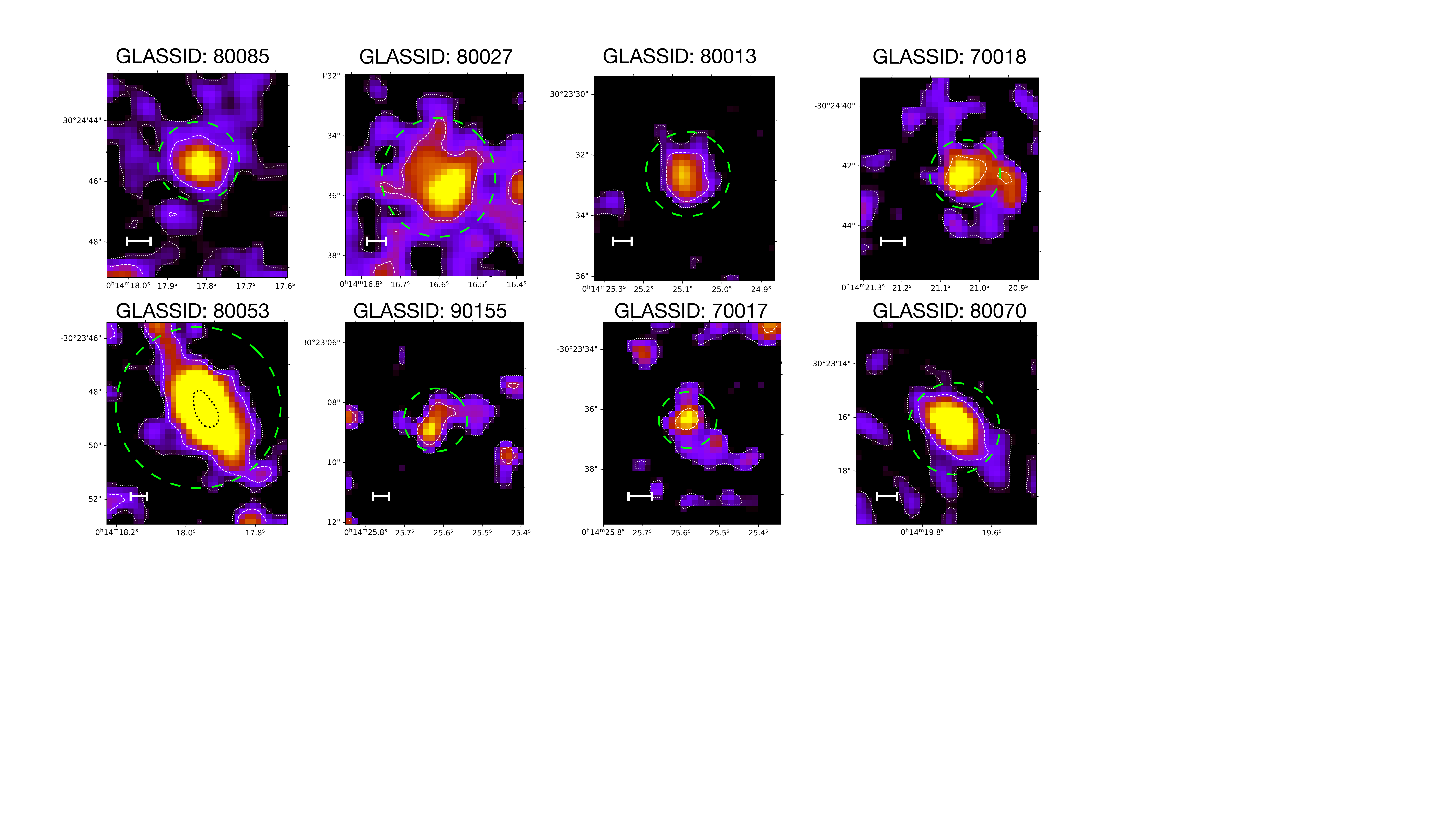}
    \caption{8 Ly$\alpha$ emitters out of the 11 objects we use for this study. Each panel shows a different object. The panels show Ly$\alpha$ narrowband image, constructed using the method described in \S\ref{lya}. Overplotted on top are the surface brightness contours at $\rm 10^{-16} \ erg \ cm^{-2} \ s^{-1} \ arcsec^{-2}$ (black dotted line), $\rm 10^{-17} \ erg \ cm^{-2} \ s^{-1} \ arcsec^{-2}$ (white dashed line), and $\rm 10^{-18} \ erg \ cm^{-2} \ s^{-1} \ arcsec^{-2}$ (white dotted line). The radius of the green dashed circle corresponds to the CoG radius (see \S\ref{linetotal}), which we used to obtain the total Ly$\alpha$ flux. The solid white line indicates 5 Kpc. }
    \label{fig:1}
\end{figure*}

\section{Data Analyses and measurements } \label{sec:method}

We construct Ly$\alpha$ narrowband images (NB) from the MUSE data, following \cite{leclercq17}, to constrain the spatial distribution of the Ly$\alpha$ emission. As later shown in Figure \ref{fig:1}, Ly$\alpha$ emission for most of our sources is diffuse and extended. Although the Ly$\alpha$ flux measurements for our sources already exist in the literature \citep{richard21}, the  values reported are severely underestimated (33-50\%). 
  Hence, to include most of the Ly$\alpha$ emission, including the diffuse low-surface brightness signal at the outskirts, we re-measured the fluxes and EWs. Details are described in the section below.  We use rest-frame optical emission lines ([OIII]$\lambda$ 5007, \Ha, \OII, \Hb) from JWST/NIRSpec spectra to measure systemic redshifts and also line ratios to derive ionization, metallicity, and dust properties. The measured \Ha \ and \Hb \  ratios enabled  dust correction of the observed line fluxes to correctly estimate the \lya \ \flya \ (see Table.~\ref{tab:1}). We start by describing the NIRSpec-derived quantities first. 

\subsection{Rest-optical line measurements} \label{opticalines}

Rest frame optical emission lines of our LAE sample are obtained with the JWST NIRSpec spectroscopy. Data reduction pipeline steps, outlined in \S\ref{jwst}, produce wavelength \& flux-calibrated, combined, rectified two-dimensional spectra for each slitlet. 1D spectrum and its associated error spectrum are extracted from the 2D spectrum using an optimal extraction algorithm \citep{horne86}. Fig.~\ref{fig:2} shows the H$\alpha$, [OIII], and [OII] lines for three example sources. 
We derive  the systemic redshift from the [OIII] $\lambda$ 5007\AA\ emission line center and constrain all the other lines to have the same redshift in the given source. The fluxes and their uncertainties were derived for each emission line by fitting a Gaussian model,  then dust corrected following the prescription outlined in \S\ref{dust} to produce the final flux ratio estimates reported in Table.~\ref{tab:1}. For calculating the ratio of Ly$\alpha$ output to the rest frame optical line flux, the optical fluxes need to be corrected by an additional slitloss fraction to compensate for the portion of the emission missed by the coverage of the NIRSpec slitless. 
The slitloss fraction is calculated as the fraction of the UV continuum light  from within the NIRSpec slitlet open shutter area to the total light from the host galaxy boundary defined by the HST segmentation map. The slitloss fraction for our sources is not large and ranges between  0.55-0.78, which is encouraging, and dictates that our line ratios are representative of most of the gas in the galaxy. 

\subsection{Dust correction} \label{dust}

We corrected the rest optical emission line fluxes for dust extinction using the Balmer decrement measurements. 
Assuming hydrogen lines emit from an optically thick HII region obeying Case B recombination, we considered the intrinsic H$\alpha$/H$\beta$ ratio  = 2.86. We adopted a \cite{calzetti00} extinction curve to compute the nebular color excess E(B-V). We took $\rm k_{H\alpha}$ = 3.33 and $\rm k_{H\beta}$ = 4.6. The corrected emission line fluxes are $\rm Line_{corrected} \ = \ Line_{obs} \times 10^{0.4.E(B-V).k_{\lambda}}$ where $\rm k_{\lambda}$ is derived from the \cite{calzetti00} extinction curve at the wavelength of the specified line.

\subsection{Ly$\alpha$ narrow-band image construction} \label{lya}

We constructed pseudo-narrowband \lya\ images using the MUSE datacube, each centered on the position and wavelength of the corresponding Ly$\alpha$ line. We closely followed the  method designed by \cite{leclercq17}. We used a wide spatial aperture to include all the detectable Ly$\alpha$ emissions above the noise level. The spectral bandwidths for constructing the \lya\ NB image were chosen to include more than 95\% of the total integrated Ly$\alpha$ line flux and to maximize the signal-to-noise within the 5$''$ $\times$ 5$''$ aperture of the said object. The chosen spectral bandwidths range between $\sim$ 10 to 30 \AA \  for our sources.

In this study, source crowding is a serious issue for many objects since almost all objects have projected close neighbors within a few arcseconds. However, these neighbors are typically at other redshifts than our chosen \lya\ emitters, so they contaminate the NB signal only with continuum emission. To properly remove the continuum, we performed median filtering in the spectral direction in a wide window of 100 spectral pixels to ignore any emission lines in the MUSE data cube \citep[similar to][]{herenz17, wisotzki16, leclercq17}. This produced a continuum-only data cube with all emission lines removed. A continuum-free data cube was then computed by subtracting this filtered data cube from the original. The resulting \lya\ images for 8 of our objects with SB contours at $\rm 10^{-16}$ (inner dotted white), $\rm 10^{-17}$ (dashed white), and $\rm 10^{-18} \ erg \ s^{-1} \ cm^{-2} \ arcsec^{-2} $ (outer dotted white) are shown in Fig.~\ref{fig:1}.

\subsection{Integrated Lyman-alpha flux and EW} \label{linetotal}

The Ly$\alpha$ flux was computed by integrating the datacubes inside the circular
aperture corresponding to a radius known as the Curve of Growth radius or ``CoG radius'' (rCoG), similar to \cite{leclercq17}. We averaged the flux in successive annuli of 1-pixel
thickness around the emission center. The COG radius was determined by the annular radius
for which the averaged flux reaches the noise value and the cumulative flux distribution flattens. The center of this last annulus corresponds to rCoG. From this aperture, we
extracted a spectrum and integrated the flux corresponding to
the Ly$\alpha$ line width; the borders of the line are set when the flux
goes under zero. This method is more robust than using a  single, fixed spatial aperture for all objects. This also ensured that most of the Ly$\alpha$ flux was included for each object. Note, due to this technique, our flux values are higher by a factor of $\sim$1.5-3 from the reported values in the public catalog of \cite{richard21}. The extracted integrated spectrum for three example objects is shown in Fig.~\ref{fig:2}. 

To calculate the EW, we needed to estimate the UV continuum at the wavelength of the Ly$\alpha$ line $f^{\rm cont}_{\rm ly\alpha}$. However, we do not detect any continuum from the MUSE spectra. Hence, we utilize HST images  mentioned in \S\ref{catalog} to estimate the UV continuum slope and derive the continuum flux at 1216\AA. These values are reported in \cite{richard21}. The observed frame EW measurement (EW$_z$) is obtained by dividing the integrated line flux by the continuum flux. The rest frame EW is then given as EW = EW$_z$ / (1+ z), where z is the redshift for the source.

\subsection{Ly$\alpha$ escape fraction} \label{lyesc}

To calculate the escape fraction, we implement a prescription similar to \cite{hayes05}, which is = $\rm L^{Ly\alpha}_{obs}  / (8.7 \times \rm \ L^{H\alpha}_{int}$ ). The case-B Ly$\alpha$/H$\alpha$ ratio is  between 8.1 and 9.0.  For consistency with previous studies, we use 8.7. Here, $\rm L^{Ly\alpha}_{obs} $ is the observed Ly$\alpha$ luminosity and $\rm L^{H\alpha}_{int}$ is the intrinsic dust-corrected H$\alpha$ luminosity. Corrected intrinsic H$\alpha$ luminosity is given by: $\rm L^{H\alpha}_{int}$ = $\rm L^{H\alpha}_{obs} \times 10. exp(0.4 . E(B-V) $ (see \S\ref{dust} for details). 

\subsection{Ly$\alpha$ FWHM and red peak velocity}
We non-parametrically characterize the full-width half maxima (FWHM) of the \textit{red component of the Ly$\alpha$ emission} using the \texttt{specutils} package of python. 
We corrected the FWHM of the Ly$\alpha$ line for the spectral
line spread function (LSF) of MUSE approximating the later
as a Gaussian. The FWHM of the LSF is wavelength
dependent, and we used the value for the MUSE-Deep Mosaic
fields (with ten hours of exposure time) given by \cite{bacon17}, which follows:
\begin{equation}
    \rm F_{mosaic}  = 5.835 \rm \times \ 10^{-8} \lambda^{-2} - 9.080 \rm \times \ 10^{-4} \lambda +5.983. 
\end{equation}
 It should be kept in mind that the FWHM measurement becomes unreliable for narrow lines, which are dominated by the LSF \citep{verhamme14, verhamme17}.
The red peak velocity of the Ly$\alpha$ profile is determined by measuring the velocity offset of the red component of the Ly$\alpha$ line relative to the systemic redshift of the sources. The systemic redshifts are determined from the brightest rest optical emission line ([OIII]$\lambda$5007\AA) from the JWST NIRSpec spectra.

\begin{figure*}
    \centering
    \includegraphics[width=\textwidth]{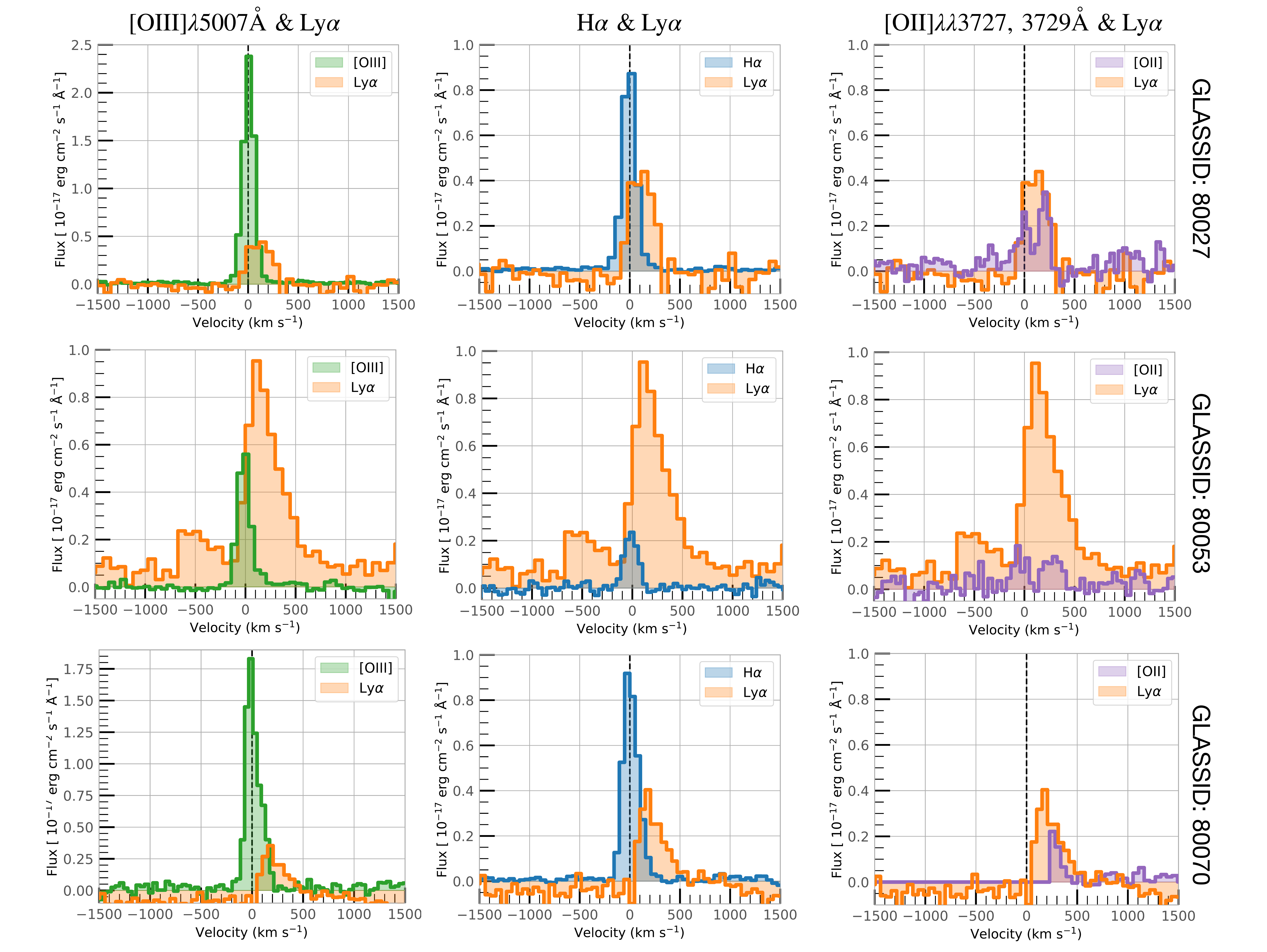}
    \caption{The Ly$\alpha$ line profiles extracted from the CoG radius from Fig.~\ref{fig:1} are shown in orange with rest-frame optical emission lines overplotted for three example LAEs in our sample. The different rows indicate different galaxies. The emission lines shown are [OIII]$\lambda$ 5007 in the first column (green), H$\alpha$ in the second column (blue), and [OII]$\lambda \lambda$3726, 3729 in the third column (purple). }
    \label{fig:2}
\end{figure*}

\section{\lya\ emission in z $>$ 3 MUSE LAE population } \label{sec4}

\begin{figure*}
    \centering
    \includegraphics[width=\textwidth]{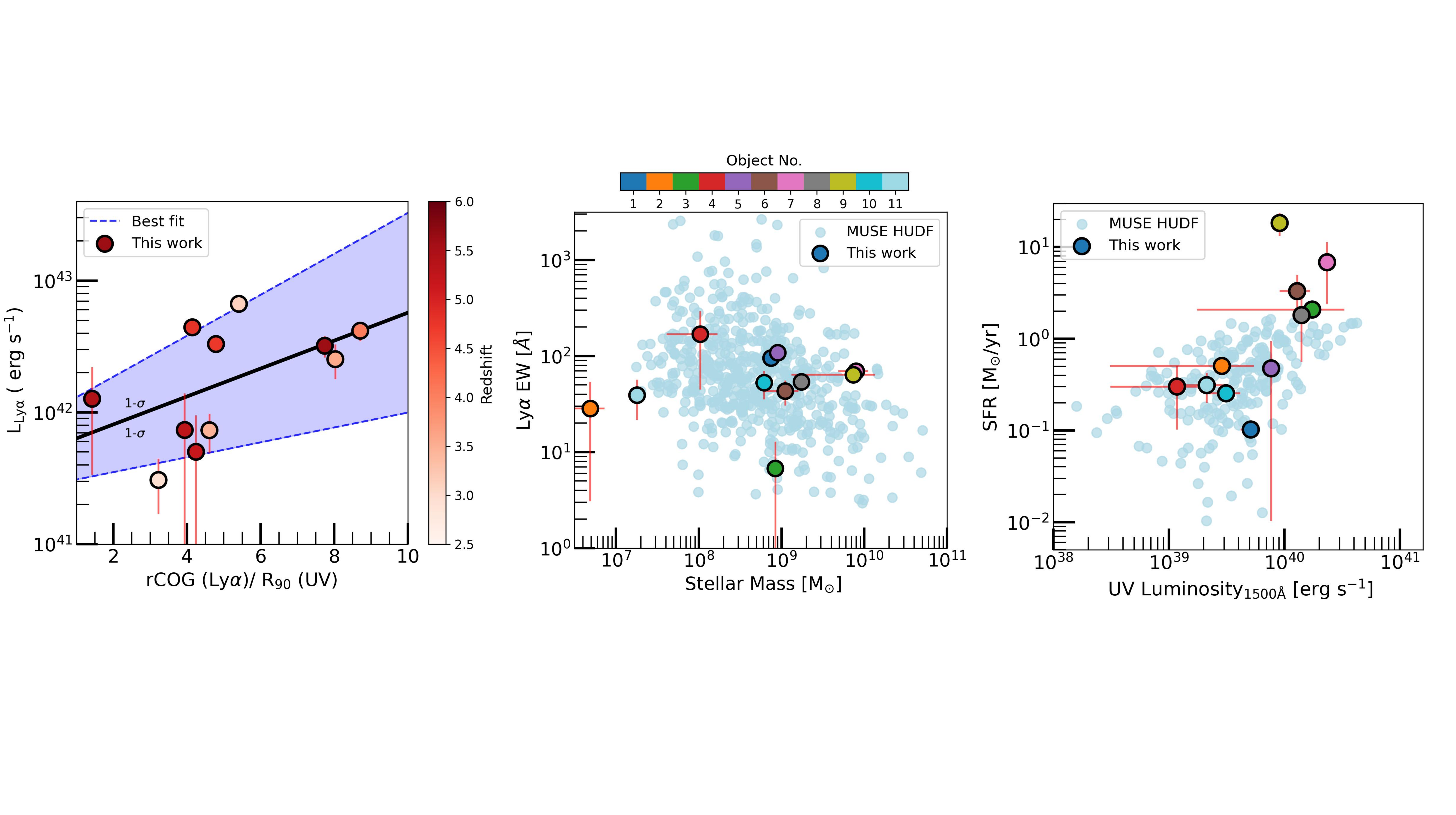}
    \caption{\textit{Left:} The total Ly$\alpha$ luminosity measured from the Ly$\alpha$ NB images (Fig.~\ref{fig:1}) as a function of the ratio of COG radius (green circles in Fig.~\ref{fig:1}) and Petrosian radius at 90\% stellar light. The black dashed line shows the best-fit relation, with 1$\sigma$ uncertainty in purple. The sources are color-coded by redshift. \textit{Middle:} Ly$\alpha$ EW vs.stellar mass for our sample of 11 LAEs (colored circles) compared against the z$\sim$ 3-6 LAEs from the MUSE Hubble Ultra Deep Field (HUDF) survey in blue \citep{bacon22}. Each of our sources is assigned a specific color (see Table.~\ref{tab:1}).  \textit{Right:} SFR vs. UV luminosity measured at rest frame 1500\AA \   wavelength for our sources (colored circles) and the MUSE HUDF sources (blue circles). Our chosen sample is largely consistent with the larger population of z$>$3 MUSE HUDF sample in terms of stellar mass, SFR, UV luminosity, and \lya\ EW. Our small number of galaxies is thus representative of z$\sim$ 3 LAEs. }
    \label{fig:3}
\end{figure*}

It is now well established that there is extended Ly$\alpha$ emission (or ``halos'') around individual LAEs \citep{wisotzki16, leclercq17}. Studies have found that the Ly$\alpha$ halo fraction goes up to 80-90\% for  UV-selected star-forming galaxies at high redshifts (2$<$z$<$5). 
Our sample is also no exception. Fig.~\ref{fig:1} shows a representative sample of eight out of eleven galaxies in our sample. The white horizontal bar in each panel shows the physical length scale of 5 kpc. The spatial extent of the \lya \ emission spans a large range, from 10 - 50 kpc. To quantify the extent of the diffuse emission, we plot the Ly$\alpha$ luminosity as a function of a distance ratio in Fig.~\ref{fig:3} (upper left panel). The latter is defined as the ratio of the COG radius indicating the spatial boundary of the Ly$\alpha$ emission (green dashed circles in Fig.~\ref{fig:1}) and the Petrosian radius of the galaxy containing 90\% of the UV continuum flux ($\rm R_{90}$). The ratio of these two measured quantities is always $>$ 1 and varies from 1.5 - 9. This indicates that the Ly$\alpha$ emission of the galaxy is more extended spatially than the host galaxy's stellar light by several factors. 
The positive trend with the Ly$\alpha$ luminosity states that galaxies with more extended Ly$\alpha$ emission have a higher COG radius and produce a higher integrated Ly$\alpha$ output. The solid black line overlaid on top shows the best-fit relation with 1$\sigma$ uncertainties indicated by the dashed blue lines. 
The extended Ly$\alpha$ emission implies that our sources have a significant amount of
cool/warm gas in the CGM.


\subsection{Ly$\alpha$ spectral shape} \label{spectral_shape}

The Ly$\alpha$ profiles of three example galaxies out of the total 11 targets in our sample are shown in Fig.~\ref{fig:2}.
We see a  variety of spectral morphologies in the \lya\ line. Most are very strong and relatively asymmetric,  with more than one component. This diversity is consistent with previous observations of galaxies with strong Ly$\alpha$ emission \citep{kulas12, erb14, erb16, henry15}. At a spectral resolution of $\rm R \sim 3000$, we see double-peaked Ly$\alpha$ profiles in three out of 11 galaxies (i.e., 27\%). Although a double-peaked profile is not ubiquitous in typical star-forming galaxies with \lya\ observations \citep{oestlin14}, they were found in 90\% of the nearby (z$<$0.3) green pea population in \cite{henry15, yang17a}. On the other hand, at higher redshift, \cite{kerutt22} found that 33\% of objects below redshift 4 have a blue peak in Ly$\alpha$ emission, but that fraction drops to 16\% for 4 $<$ z $<$ 5. The significant drop in the blue peak fraction at higher redshift is due to the rising neutral gas fraction in the intervening IGM, eating away the blue component and leaving only the red peak behind in the observed spectra.  These are consistent with our high redshift sample at z = 3-6.


\subsection{Is our sample representative of the larger 3$<$ z $<$ 6 MUSE LAE population? } \label{compare_with_MUDF}

Our primary goal in this study is to measure any correlation between \flya \ and other physical properties facilitating \flya\ for our z=3-6 LAE sample and to determine their similarity with the low redshift analog galaxies. First, we test whether our sample is representative of the 
larger population of high redshift LAEs in terms of Ly$\alpha$ output 
and host galaxy properties. 
We compare the Ly$\alpha$ EW vs. stellar mass of our sample (colored circles; each color indicating a source in Table.~\ref{tab:1}) to the  MUSE z$\sim$ 3-
6 LAEs (blue circles) drawn from the MUSE HUDF dataset \citep{bacon22}, mentioned in 
\S\ref{comparisonsample}. As shown in Fig.~\ref{fig:3} (middle panel), galaxies in our 
sample show the stellar mass and Ly$\alpha$ EW distribution to be extremely consistent 
with the HUDF sample while covering a large range in stellar mass. There is also a clear 
negative correlation between the two quantities. Indeed, most observational studies of 
the stellar populations of Ly$\alpha$ galaxies suggest that the LAEs favor relatively 
young and low-mass systems \citep{ono10, hayes23}, although see \cite{finkelstein09} for 
counterexamples. This phenomenon occurs due to the 
increased gas and dust content of more massive galaxies, which makes Ly$\alpha$ escape 
difficult, thereby decreasing the measured Ly$\alpha$ EW. 9 out of  11 galaxies in our 
sample have stellar masses below $\rm 10^{9.5} \ M_{\odot}$. Note that two galaxies in our sample have extremely low values of stellar mass with $\rm M_{\star} < 10^{7.3}  \ M_{\odot}$. These two galaxies fall outside the range of the MUSE HUDF LAEs with well-constrained masses but may be consistent with the population of HUDF \lya\ sources with little to no detected continuum in the HST imaging \citep{maseda17}.  %

In Fig.~\ref{fig:3} (right panel), we show the star formation rate vs. the UV continuum luminosity measured at 1500\AA.  Our sources occupy a similar region in the parameter space as the MUSE sample. Three sources show SFR $> \rm 2 
 \ M_{\odot} / yr$, slightly higher than the typical values in the HUDF sample. This possibly arises from the different star formation rate estimators used in different catalog measurements. The positive correlation, seen in the figure, is expected since the SFR is derived from the UV luminosity estimates. The star formation rates of our sample vary between $\sim \rm 0.1 - 10 \ M_{\odot}/yr$, which indicates a diverse population but representative of high z LAEs previously reported in the literature. 

\section{  How do the low redshift analog galaxies compare with galaxies at redshift $>$ 3? }

With the measurements of \lya\ emission properties and the rest optical emission line ratios, we are ready to investigate how our high z LAE sample compares with the low redshift analogs and LyC emitters \citep{hayes23, flury22}.

\subsection{Ly$\alpha$ FWHM vs. red peak velocity}


With the availability of the JWST NIRspec spectra, we can measure the systemic redshift of our z$>$ 3 sources based on the brightest rest optical emission lines. This enables us to determine the Ly$\alpha$ red peak velocity offset with confidence. 
In Fig.~\ref{fig:4} in the top left panel, we show the FWHM of the Ly$\alpha$ profiles as a function of the velocity offset for our targets. Our 3$<$ z $<$ 6 LAE sources are shown in circles, with each color indicating a different source, as listed in Table.~\ref{tab:1}. The blue squares indicate the z$<$0.5 galaxies, which are analogs to the high z LAEs \citep[sample from ][]{hayes23}.  
We find a strong positive correlation between the velocity and FWHM in our sample, echoing the trend observed with the low redshift analog galaxies. This is also in strong agreement with the prediction from the radiative transfer (RT) theory, which states that low, neutral hydrogen column density causes Ly$\alpha$ photons to scatter less and thus helps them to escape more easily. This shapes the Ly$\alpha$ line to be narrower (low FWHM), brighter (enhanced flux), and less redshifted (low-velocity offset) compared to the systemic velocity.

This similarity between the low z and high z samples and their agreement with the RT models suggests that the mechanisms regulating the output of \lya\  may not vary with redshift.  
The mean red peak velocity for our sample is $ \rm 207 \pm  53 \  km \ s^{-1}$. This is consistent with the measurements from the low-z analogs, with a mean velocity = 241 $\rm km \ s^{-1}$. This strong positive correlation for z$>$ 3 galaxies indicates that these properties may be related intrinsically in $z>6$ galaxies as well, although both are impacted by neutral IGM gas at these redshifts.
This empirical relation can be used to derive the systemic redshifts of even higher redshift galaxies from the measurement of the FWHM of the Ly$\alpha$ line alone.

\subsection{Ly$\alpha$ escape fraction}

\begin{figure*}
    \centering
    \includegraphics[width=\textwidth]{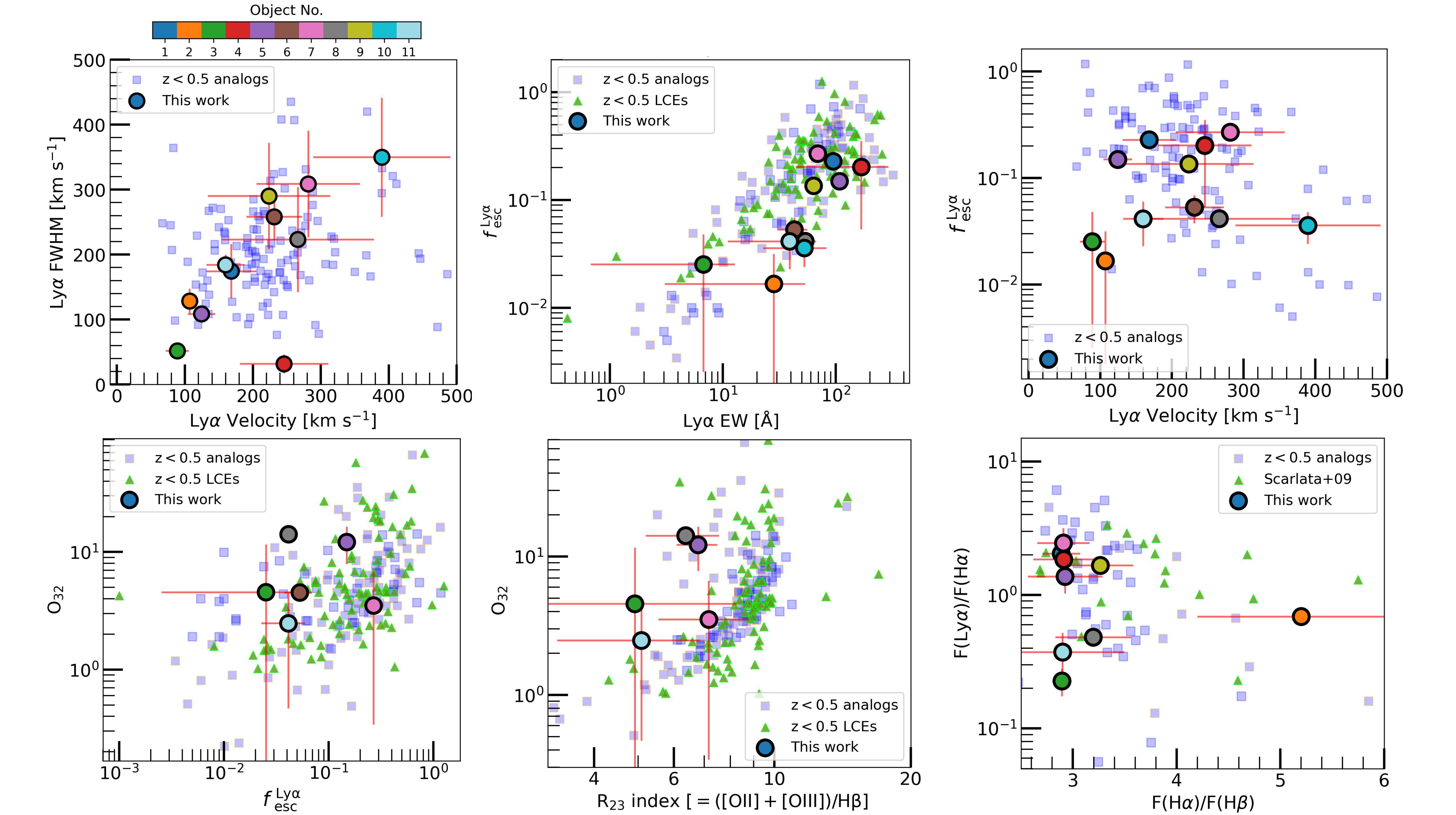}
    \caption{\textit{Upper left:} FWHM of the red peak of Ly$\alpha$  vs. red peak velocity offset for all 11 sources in our study (colored circles) compared to the low-z analog sample from \cite{hayes23} (blue squares). The velocity offsets are calculated with respect to the systemic velocity measured from the JWST/NIRSpec spectra.  Each of our objects is assigned a specific color, as indicated in the color scale. The object identifier numbers correspond to the numbers quoted in the first column of Table.~\ref{tab:1}. \textit{Upper middle:}  Ly$\alpha$ escape fraction vs. Ly$\alpha$ EW for our sources (circles) compared with the low z analog galaxies \citep{hayes23} (blue squares) and low redshift Lyman continuum emitters from the LzLCs survey \citep{flury22} (green triangles). \textit{Upper right:} Ly$\alpha$ escape fraction vs. velocity offset. Symbols and colors are the same as in the previous panel. 
    \textit{Lower left:} $\rm O_{32}$ ratio vs. Ly$\alpha$ escape fraction for our sources with detected [OIII] \textit{ and } [OII] emission lines (circle). Low z analog galaxies \citep{hayes23} and low z Lyman continuum emitters \citep{flury22} are shown in blue squares and green triangles, respectively. 
      \textit{Lower middle:} 
      $\rm O_{32}$ ratio vs. $\rm R_{23}$ index, which is defined as ([OII] + [OIII])/H$\beta$. $\rm R_{23}$ is a metallicity indicator. Symbols and comparison samples are the same as in the middle panel.   \textit{Lower right:} Ly$\alpha$/H$\alpha$ vs. H$\alpha$/H$\beta$ ratio for our sample (red circles) compared with low z analogs \citep[blue squares; ]{hayes23} and GALEX grism selected dusty LAEs \citep[green triangles; ][]{scarlata09}.   See text for details.}
    \label{fig:4}
\end{figure*}

We estimate the escape fraction of Ly$\alpha$ photons reaching the
detector from the Ly$\alpha$/H$\alpha$ flux ratio, as described in \S\ref{lyesc}. We find that the escape fraction varies between 0.02 - 0.26, with a mean \flya \ = 0.10$\pm$0.03, which is slightly lower than the low redshift  population (mean \flya \ = 0.23). High redshift LAEs from existing studies find a large range of escape fraction: \cite{hayes10} find \flya\ $\rm \sim  0.05$ while \cite{steidel11} reports \flya \ $\rm \sim$ 0.3 for z$\sim $ 2 LAEs. \cite{trainor15} also estimate an
escape fraction of $\sim$ 30\% for a sample of faint LAEs at z $\sim$ 2.7. However, we should note that many of these previously existing studies of z$>$ 2 LAEs lacked rest-optical spectra with sufficient sensitivity to measure H$\alpha$ flux and correctly perform dust correction.   
The \flya\ values for our sample are listed in Table.~\ref{tab:1} and are plotted against Ly$\alpha$ EW in the upper middle panel of Fig.~\ref{fig:4} (circles). The low redshift analogs and the LzLCs samples are also shown using squares and triangles, respectively. 
The two quantities show a positive correlation for our sources and are consistent with the values seen in the low redshift LAEs. Five of our galaxies show a Ly$\alpha$ escape fraction $>$10\%. 
Interestingly, the low-redshift samples show some spread in \lya\ EW for a given \flya, which could be related to the stellar population properties. Multiple factors like initial mass function, metallicity, or the presence of binaries can make the ionizing spectrum to be harder, 
which  can produce more ionizing photons, and therefore Ly$\alpha$ photons, for a given non-ionizing UV luminosity \citep{malhotra02}. Compared to the low-z analogs, our sample falls more towards the higher EWs for a given \flya, which could indicate more extreme stellar populations in the $z>3$ galaxies.  

In general, when the column density of the gas is low, Ly$\alpha$ photons scatter much less - resulting in
smaller velocity offsets and greater Ly$\alpha$ escape. 
A similar relationship is also seen in the Lyman continuum leakers. Lower velocity offsets exhibit broader lines, a greater Lyman continuum escape, and a greater Ly$\alpha$ escape fraction \citep{izotov18}. Thus, we expect a negative trend between the velocity offset and the Ly$\alpha$ escape fraction. We show the relation between these two quantities for our objects (circles) in the upper right panel in Fig.~\ref{fig:4}, compared with the low redshift analog galaxies (squares). Taken alone, our sample does not show any trend, and  there is a considerable scatter in our high z sources. A larger sample can provide better inference on the validity of this relation for high redshift galaxies.


\subsection{Galaxy morphology}

The concentrated star formation and high SFR surface densities are necessary for producing an ample amount of ionizing radiation, thus, creating excess Ly$\alpha$ photons. Hence, compact, highly star-forming galaxies host strong Ly$\alpha$ emissions. Indeed, our sample of z$>$3 LAEs is sufficiently compact in UV continuum size, as observed from the HST/ACS images. Our sample's mean effective radius ($\rm R_e$) is $\sim$ 1.1 kpc, with some galaxies exhibiting $\rm R_e$ as low as 0.1 kpc. This is expected since galaxies with strong Ly$\alpha$ emission likely represent galaxies in earlier stages of evolution with younger ages and smaller sizes. The effective sizes of our sources are given in Table,~\ref{tab:1}.
 This finding is also qualitatively consistent with that of \cite{malhotra12}, and \cite{hayes23}. They found that LAEs, in general, are drawn from the more compact end of the size distribution of normal star-forming galaxies. Quantitatively, \cite{malhotra12} find LAEs to have half-light radii close to 1.0 kpc at
redshifts 2–6, which is in remarkable agreement with our measurement of the effective radius.

\subsection{Variation of [OIII]/[OII] ratio with Ly$\alpha$ escape fraction}

The [OIII]$\lambda$5007\AA  /[OII]$\lambda$3727,29 \AA\  ratio (or $\rm O_{32}$, in short) measures the ionization state of the gas in a galaxy. 
A classical HII region model uses two zones. If the neutral gas gets sufficiently depleted due to the ionizing photons from young stars, the outer edge of the HII regions gets truncated. This reduces the [OII] emission, resulting in a high $\rm O_{32}$ ratio. This scenario indicates low, neutral gas column density channels that allow Ly$\alpha$ and Lyman continuum photons to escape more easily. Thus a high $\rm O_{32}$ ratio has been proposed to trace ``density-bounded'' neutral gas regions in \lya \ emitting galaxies, which can generate a lot of ionizing radiation, and thus a lot of Ly$\alpha$ photons \citep{jaskot13, nakajima14}.

\cite{izotov16} found observational evidence for the first time that high $\rm O_{32}$ ratios are a potential signature of Lyman continuum leakage and can trace a low-density path through the interstellar medium of galaxies along the line of sight. Similarly, \cite{nakajima13, nakajima14} found that galaxies with a high
$\rm O_{32}$ ratio in low redshift analog populations like the green peas, Lyman break analogs, and in high redshift z = 2–3 galaxies have a high Ly$\alpha$ escape fraction.
Figure.~\ref{fig:4} lower left panel shows the $\rm O_{32}$ ratio as a function of the Ly$\alpha$ escape fraction for our sample (circles). We show the Low redshift Lyman continuum survey (LzLCS) galaxies from \cite{flury22} (green triangles), and low redshift analog galaxies from \cite{hayes23} (blue squares) for comparison. The LzLCs sample and the low z analog galaxies show a positive trend between the two quantities. 
In our sample, only 6 out of 11 galaxies have simultaneous detection of both [OIII] and [OII] emission lines to derive the $\rm O_{32}$ ratio. Our sources have $\rm O_{32}$ ratio between 2.5-14.1 and are overall consistent with the positive correlation between [OIII]/[OII] and \flya\ seen in the low redshift analog galaxies. 
Establishing this correlation at high redshift is crucial since $\rm O_{32}$ is one of the key diagnostics available to JWST at $z>7$.
 


\subsection{Ionization and metallicity}

Gas-phase metallicity is a key property of the host galaxy ISM since it is a record of
a galaxy’s star formation history and gas infall/outflow. Metallicity
estimates can be made with metal lines divided by hydrogen recombination lines, such as ([O III] $\lambda \lambda$ 5007, 4959 + [O II] $\lambda$3727)/H$\beta$ \citep[$\rm R_{23} index; $][]{pagel79, kewley02}.
\cite{cowie11} found that low redshift LAEs exhibit lower metallicities, compact sizes, and younger ages  compared to a UV-continuum-selected sample of star-forming galaxies at similar stellar mass and redshift. These findings are consistent with the idea that LAEs are galaxies in the early stages of evolution. 
On the other hand, the ionization state, traced by $\rm O_{32}$, is sensitive to the degree of excitation and the optical depth of the
HII region in a galaxy \citep[e.g., ][]{brinchmann08}; a large $\rm O_{32}$ may be due to a low optical depth and a high escape fraction of ionizing photons, as discussed in the previous section. \cite{nakajima13, nakajima14}, for the first time, presented the ionization and metal properties  for a very small sample of z$\sim$ 2 LAEs based on multiple nebular emission lines. They found that the high z LAEs have a higher typical ionization parameter and lower metallicity in contrast with other high z galaxies. 



We show the relationship between the $\r O_{32}$ and  the $\rm R_{23}$ index (Fig.~\ref{fig:4} lower middle panel) for our sources in circles, compared with the low redshift analog population: \cite{hayes23} (blue squares) and low redshift Lyman Continuum emitters \cite{flury22} (green triangles). Our sources occupy a broad region in the $\rm O_{32}$ vs. $\rm R_{23}$ parameter space, although the uncertainty of the measurements is also considerably high. This large scatter in the $\rm O_{32}$ vs. $\rm R_{23}$ diagram has also been seen with other \lya \ emitters at z $>$ 4 using JWST measurements \citep{mascia23}. This could imply that high redshift LAEs exhibit a vast range of metallicity and ionization states of the gas, or it could simply be an effect of large uncertainties in the measurements.  

\subsection{Dust extinction and reddening}

The effect of dust content on Ly$\alpha$ emission has been
studied extensively \citep{scarlata09, finkelstein11, hayes10, cowie11, nakajima12, henry15}. 
Ly$\alpha$ photons produced from the star-forming regions can undergo multiple scattering from the HI gas in the ISM and CGM of galaxies. With the increase in the number of scattering, the probability of the Ly$\alpha$ photons to be absorbed by dust grains also increases. Thus, the observed line intensities depend on the number of ionizing photons and the attenuation produced by dust along the line of sight. High dust content indicates low \lya \ output. 

In Fig.~\ref{fig:4} lower right panel, we show the Ly$\alpha$/H$\alpha$ flux ratio as a function of the observed H$\alpha$/H$\beta$ ratio for our objects. Comparing our sample with other Ly$\alpha$ emitting galaxies with published Ly$\alpha$ and optical line ratio measurements is crucial. We separate the low z analog sample into two groups -- galaxies identified as ``dusty'' LAEs from GALEX grism surveys \citep{scarlata09} shown in green triangles, and the rest of the low redshift analog galaxies from \cite{hayes23}, shown in blue, that include galaxies from the Ly$\alpha$ reference survey \citep{hayes13, hayes14}, green pea population \citep{yang17a}, blueberries \citep{yang17b} and Lyman Break Analogs \citep{heckman15}. 
In the absence of dust, for case B recombination at $\rm T_e \sim 10^4 \ K$ and $\rm n_e = 10^2 \ cm^{-3}$, the expected  line ratios are 2.86 and 8.7 for H$\alpha$/H$\beta$ and Ly$\alpha$/H$\beta$ ratios respectively \citep{pengelly64}. 
Galaxies in our sample show a large range in $\rm Ly\alpha/H\alpha$, but a comparatively small range in   H$\alpha$/H$\beta$. The mean H$\alpha$/H$\beta$ ratio is 3.2$\pm$1.3, indicating low  dust content. This is similar to the low z analog galaxies, where 90\% of the population has H$\alpha$/H$\beta$ $<$ 4. 9 out of 11 sources have simultaneous detection of both H$\alpha$ and H$\beta$, while the other two sources miss H$\beta$ line due to detector gaps. 8 out of those 9 galaxies have  H$\alpha$/H$\beta$ $<$3.3 and are greatly consistent with the low redshift analogs, particularly the Green pea population \citep{henry15, yang17a}. One galaxy shows a very high dust content with a value of 5.2, which is more consistent with the 
local dusty LAEs of \cite{scarlata09}, shown in Figure. The low dust content for the LAEs is consistent with the prediction that a higher amount of dust content scatter and absorbs the \lya \ photons more, making their escape difficult. The Ly$\alpha$/H$\alpha$ ratio for all our sources is smaller than the value of 8 - 9 predicted from case B. They vary by almost an order of magnitude. 7 of our high z LAE sources show Ly$\alpha$/H$\alpha$ overall consistent with the low redshift population. The remaining two sources have considerably low Ly$\alpha$ output.

\begin{table*}[!htb]
\caption{Ly$\alpha$ and rest-optical properties of $z\sim3-6$ galaxies in the GLASS-JWST program used in this work.}\label{tab:1}
\centering
\setlength\tabcolsep{1.6pt}
\begin{tabular}{||l|| c c c c c c c c c c c c||}
\hline
No. & ID & $\rm R_{e}$ & $z_\mathrm{sys}$ & $\rm V_{Ly\alpha}$  & $\rm FWHM_{Ly\alpha}$  & $\rm F_{Ly\alpha}$   & Ly$\alpha$ EW  &   \flya \   & $\rm O_{32}$* & F(Ly$\alpha$)/F(H$\alpha$)** & F(H$\alpha$)/F(H$\beta$)** & $R_{23}$* \\
 & & (kpc) & & (km/s) & (km/s) & ($10^{-20} \ \rm erg \ cm^{-2} \ s^{-1}$) & (\AA) & & & & &  \\
\hline
\hline
1 & 70003 & 0.12 & 5.618 & 168 $\pm$ 37 & 174 $\pm$ 41 & 941.1 $\pm$ 173.7 & 95 $\pm$ 17 & 0.23 $\pm$ 0.04 & -- & 2.0 $\pm$ 0.4 & 2.89 $\pm$ 0.18 & -- \\ 
2 & 70017 & 0.33 & 5.186 & 107 $\pm$ 5 & 128 $\pm$ 19 & 177.7 $\pm$ 158.5 & 28 $\pm$ 25 & 0.02 $\pm$ 0.01 & -- & 0.7 $\pm$ 0.0 & 5.2 $\pm$ 1.18 & -- \\ 
3 & 70018 & 0.77 & 5.282 & 89 $\pm$ 17 & 51 $\pm$ 10 & 249.2 $\pm$ 224.1 & 6 $\pm$ 6 & 0.03 $\pm$ 0.02 & 4.5 $\pm$ 6.0 & 0.2 $\pm$ 0.1 & 2.89 $\pm$ 0.09 & 4.9 $\pm$ 3.3 \\ 
4 & 70022 & 3.34 & 5.429 & 245 $\pm$ 65 & 31 $\pm$ 14 & 403.5 $\pm$ 296.7 & 168 $\pm$ 123 & 0.2 $\pm$ 0.15 & -- & 1.8 $\pm$ 0.5 & 2.91 $\pm$ 0.29 & -- \\ 
5 & 80013 & 0.7 & 4.043 & 124 $\pm$ 20 & 108 $\pm$ 12 & 2653.7 $\pm$ 444.1 & 108 $\pm$ 18 & 0.15 $\pm$ 0.02 & 12.1 $\pm$ 4.3 & 1.4 $\pm$ 0.3 & 2.92 $\pm$ 0.36 & 6.8 $\pm$ 0.7 \\ 
6 & 80027 & 0.92 & 3.58 & 231 $\pm$ 41 & 257 $\pm$ 36 & 2154.8 $\pm$ 632.4 & 43 $\pm$ 12 & 0.05 $\pm$ 0.02 & 4.5 $\pm$ 0.2 & 0.6 $\pm$ 0.1 & -- & -- \\ 
7 & 80053 & 1.55 & 3.129 & 281 $\pm$ 76 & 308 $\pm$ 82 & 7829.7 $\pm$ 705.2 & 69 $\pm$ 6 & 0.27 $\pm$ 0.02 & 3.5 $\pm$ 3.2 & 2.4 $\pm$ 0.7 & 2.91 $\pm$ 0.25 & 7.2 $\pm$ 1.6 \\ 
8 & 80070 & 0.9 & 4.797 & 266 $\pm$ 112 & 223 $\pm$ 80 & 1881.0 $\pm$ 236.8 & 54 $\pm$ 9 & 0.04 $\pm$ 0.01 & 14.1 $\pm$ 0.8 & 0.5 $\pm$ 0.1 & 3.2 $\pm$ 0.37 & 6.4 $\pm$ 1.1 \\ 
9 & 80085 & 0.8 & 4.725 & 223 $\pm$ 90 & 289 $\pm$ 81 & 1454.9 $\pm$ 193.8 & 63 $\pm$ 8 & 0.14 $\pm$ 0.02 & -- & 1.7 $\pm$ 0.2 & 3.26 $\pm$ 0.32 & -- \\ 
10 & 80113 & 0.95 & 3.472 & 389$\pm$ 101 & 349 $\pm$ 91 & 669.6 $\pm$ 221.3 & 53 $\pm$ 17 & 0.04 $\pm$ 0.01 & -- & 0.4 $\pm$ 0.1 & -- & -- \\ 
11 & 90155 & 0.7 & 2.94 & 159 $\pm$ 28 & 184 $\pm$ 15 & 418.0 $\pm$ 187.1 & 39 $\pm$ 18 & 0.04 $\pm$ 0.02 & 2.5 $\pm$ 2.0 & 0.4 $\pm$ 0.1 & 2.9 $\pm$ 0.59 & 5.1 $\pm$ 1.8 \\
\hline

\hline
\end{tabular}
\tablecomments{ No. indicates Object Serial number matching the color scale in Fig.~\ref{fig:3} \& \ref{fig:4}. * $\rm O_{32}$ and $\rm R_{23}$ denotes the dust corrected ratio of ($\rm [OIII] \lambda$5007\AA/[OII] $\rm \lambda$3727, 29\AA) and ($ \rm [OIII] \lambda 5007$ + $\rm [OII] \lambda 3727, 39$) / H$\beta$ respectively. For sources where H$\beta$ falls in the detector gaps, we perform dust correction assuming a mean E(B-v) = 0.09. **Observed flux ratios.  }
\end{table*}

\section{Discussion} \label{discussions}

\subsection{Comparison with low z analogs and LyC leakers}

\lya \  is possibly the best indirect diagnostic of LyC escape since the conditions that favor the escape of \lya \  photons are often the same that allow for the escape of LyC photons.  
In this work, we concentrate on studying \lya \  emission for galaxies with z $<$ 6, thus avoiding the significant IGM attenuation but being close enough in redshift to the EoR. We study the correlations between \flya  \ with another indirect but promising diagnostic tested at low redshift by \citep{flury22}: $\rm O_{32}$.  One of our primary goals is to determine whether the correlations observed in the low-z population prevail in the z = 3 - 6 regime as well. We compare our sample with two main low z samples – 1) low z ``analogs'' from \cite{hayes23} - which include green peas, blueberries, Lyman break analogs, intense starbursts, and extreme emission line galaxies, and 2) low redshift Lyman continuum emitters from \cite{flury22}, which are candidates for LyC leakers. 

In Fig.~\ref{fig:4}, we plot the relation between \flya\  with \lya\  EW, velocity offset, and $\rm O_{32}$ for our JWST high redshift sample and compare them with the local population  \citep{hayes23, flury22}. A huge caveat for our study is  the small size of our sample. The objects in our sample do not show any strong correlations by themselves. Here we discuss if they are generally consistent with the correlation parameter space occupied by the low redshift galaxies. 

We find that the \flya \  vs. \lya \  EW generally shows a positive trend, as expected. However, two sources (Object 8, 10) show lower \flya\  than expected for the given \lya \ output. One of them (Object 8) shows a higher dust content which might contribute to the lower \flya. We next analyze the \flya\  vs velocity offset. 
Our sources are consistent with the \cite{hayes23}  low-redshift analogs, although there is a significant scatter. Our systemic redshift measurements are determined from the brightest rest optical lines (\Ha \ and [OIII]5007\AA) measured from the NIRSpec spectra. Uncertainties in wavelength calibration in JWST/NIRSpec and MUSE spectra can contribute to this scatter. 
However, two sources (Object 2 and 3) have the lowest \flya \  and \lya\  velocity and hence lie completely offset from the low z analogs. These are also two of our sample's highest redshift galaxies (z = 5.186 and 5.282). We hypothesize that this could be caused by IGM attenuation. 

We now focus on $\rm O_{32}$. 
High $\rm O_{32}$ has been proposed as an indicator of higher \fesc\  \citep[e.g., ][]{nakajima14}. The reasoning is that the high $\rm O_{32}$ ratio selects highly ionized systems, which are more likely to have density-bounded channels through which ionizing photons can escape. 
We do not have a direct estimate of LyC \fesc, so we plot $\rm O_{32}$ vs. \flya. We find that all 6 sources of our sample with measured $\rm O_{32}$ indeed mostly lies in the region of the plot populated by low-redshift LyC leakers \citep{flury22} and analogs sample \citep{hayes23}. The median $\rm O_{32}$ is 4.5 $\pm$ 1.2. We see that the majority of our sources (4 out of 6 galaxies) show $\rm O_{32} < $ 5, which has been indicated as a lower threshold for LyC leakers with an \fesc $>$ 0.05 \citep{flury22}. Thus our objects are predicted to have \fesc $<$ 0.05. Although, some studies have shown that $\rm O_{32}$ does not necessarily correlate well. with \fesc \citep{naidu18, katz20}, and thus are not expected to correlate with \flya \  as well; nonetheless, samples of LyC leaking galaxies at low-redshift generally show that the fraction of galaxies with high \fesc increase toward higher O32, even if the correlation is not tight \citep[e.g., ][]{izotov16, flury22}. This is precisely what we find in our sample with \flya\ as well. 

Finally, the $\rm O_{32}$ vs $\rm R_{23}$ index diagram is widely used to examine the gas-phase metallicity and ionization state both in the local universe \citep[e.g., ][]{izotov16, izotov18} and at high redshift \citep{flury22, nakajima20, reddy22, vanzella19}. Recent studies by \cite{nakajima20} showed that z $\sim$ 3 LyC leakers tend to populate the upper right part of this diagram, i.e., they have high $\rm O_{32}$ and high $\rm R_{23}$. This result is also seen in high-resolution  cosmological radiation hydrodynamics simulations \citep{katz20}. However, both these studies conclude that the $\rm O_{32}$ vs $\rm R_{23}$ plane is not the most useful to differentiate between leakers and non-leakers. Our sample occupies a broad region in the parameter space, which could reflect either a wider range of metallicity and ionization states or the fact that we have a large measurement uncertainty, very similar to 4.5 $<$ z $<$ 8 \lya \  emitting galaxies studied by \citep{mascia23}. 

\subsection{Prediction of LyC escape fractions} 

We try to indirectly estimate \fesc\  from our measured properties. Previous studies have attempted to estimate \fesc\  based on neutral gas properties and low ionization absrobtion lines. It is difficult to detect such lines and obtain neutral gas properties at higher redshifts. Hence, we adopt the fully-data driven regression analysis on observable galaxy properties, presented by \cite{mascia23}, for a similar sample of z = 4-8 \lya\  emitting galaxies in the GLASS dataset. Note that
\cite{mascia23} used the low redshift Lyman continuum
survey to derive this relation and then applied it to z = 4-8
LAEs.The best-fit relation they proposed is:

\begin{equation}
    \rm log_{10}(f_{esc}) = A + B *log_{10}(O_{32}) + C*r_e + D*\beta
\end{equation}

Where A = -1.92, B = 0.48, C = -0.96, D = -0.41, $\rm r_e$ is the effective radius in kpc (Table.~\ref{tab:1}) and $\beta$ is the mean UV slope = -2.3$\pm$0.4, calculated by Prieto-Lyon et al. 2023 (in prep). We find that the \fesc\ varies between 0.03 - 0.07 with a mean value = 0.04. This is consistent with what we previously hypothesized – the majority of our sources should have \fesc $<$ 0.05 based on the $\rm O_{32}$ ratio alone. 
The main limitation of our study is the small sample size. Our results are mostly based on 6 sources with the detection of both [OIII] and [OII] lines. A larger and more evenly distributed sample would be required to draw any conclusion. Still, our \fesc\ is rather low, with the average value lower than the median \fesc$\sim$0.1 predicted by \cite{naidu20} and \cite{mascia23} for galaxies with median z = 6.

\section{Summary}  \label{summary}

Thanks to the deep spectroscopic data obtained with MUSE  and the outstanding JWST NIRSpec spectroscopy on selected targets in the  field of the Abell 2744 cluster: we have been able to study the rest frame UV-optical spectral properties of 11 Ly$\alpha$ emitting galaxies with spectroscopic redshift in the range 3 $<$ z $<$6.  We aimed to answer how the  \lya\ emission and their correlations with the galaxy properties compare between our high redshift sample with the existing low redshift analog population. Using the combined analyses of JWST NIRSpec and MUSE spectra, we were able to measure accurate estimates of systemic redshifts, \lya\ velocity offsets, \flya, ionization states, metallicity indicators, and dust content. We report the measurements of the most prominent rest optical emission lines (H$\alpha$, [OIII], [OII], H$\beta$).  Our main results are summarized as follows:

\begin{enumerate}
    \item All 11 galaxies in our sample are detected to have diffuse extended Ly$\alpha$ emission. We find a large range of physical sizes of these extended Ly$\alpha$ emissions ranging from 10 - 50 kpc, often extending beyond the host galaxy's stellar component  \citep[similar to ][]{leclercq17, hayes13}. Our Ly$\alpha$ spectral profiles show a diverse morphology, with 3/11 showing a double-peaked profile.

    \item The full-width half maxima of the Ly$\alpha$ emission shows positive correlations with the \lya\ red peak velocity offset. Thus, the low-z derived relation works well for z$<$ 6 galaxies and can be used to estimate systemic redshifts based on measurements of \lya \ FWHM alone, as previously proposed by \cite{verhamme17}. 

    \item We compared our 11 galaxies to the low redshift LyC emitters from \cite{flury22} and low redshift analogs from \cite{hayes23}. Although our galaxies do not show many strong correlations if taken alone, their properties are entirely consistent with the low redshift population in terms of \lya \ EW, \flya, $\rm O_{32}$, \lya/\Ha \  and dust content.

    \item Our high z LAE sample has low  dust content (all except three have  \Ha/\Hb $<$ 3 ), compact sizes (mean $\rm R_{e} \sim $ 1~kpc ), low mass ($\rm M_{\star} < 10^{10} \ M_{\odot}$) and high SFR (SFR $\sim$ 0.1 - 10 $\rm M_{\odot}/yr$). These are consistent with the prediction from radiative transfer models and imply that Ly$\alpha$ emitting galaxies define the early stages of a galaxy's lifecycle. The low dust allows the \lya \ photons to escape without being completely absorbed. The striking similarity of the \lya/\Ha \ ratio vs. Balmer decrement with the low redshift analog population is remarkably evident even for our small sample of sources. 

    \item  We use the empirical relation  between LyC escape fraction and the
three galaxy parameters proposed by \cite{mascia23}, to predict LyC \fesc. We find that our sources are not strong LyC leakers, with an average LyC escape fraction $\sim$ 0.04.

    In conclusion, our low mass high redshift galaxies have physical and spectroscopic properties to be broadly consistent with the low redshift population, which are rightly considered ``analogs'' of high redshift LAEs. This suggests that diagnostic relations for \lya\  and LyC, derived using low-z analogs \citep{flury22, runnholm20}, may be applicable in early epochs. With the caveat of a small sample size, we found that the amount of escaping ionizing photons is not large in our galaxies (\fesc $\sim$ 0.03 - 0.07).
    A larger sample covering a wider range in redshift, stellar mass, and SFR is needed to make a stronger statement about the nature of correlations seen in the high z LAEs. Larger JWST samples taken, for example, from the JWST JADES GTO program, will be suitable for studying more high redshift LAEs and better calibrating the correlations we show here.

\end{enumerate}


\begin{acknowledgements}

This work is based on observations made with the
NASA/ESA/CSA \textit{James Webb Space Telescope}. The data
were obtained from the Mikulski Archive for Space Telescopes at the Space Telescope Science Institute, which is operated by the Association of Universities for Research in Astronomy, Inc., under NASA contract NAS 5-03127 for \textit{JWST}. These observations are associated with program JWST-ERS1324. The \textit{JWST} data used in this paper can be found on MAST: http://dx.doi.org/10.17909/fqaq-p393. We acknowledge financial support from NASA through grant JWST-ERS-1342.
The authors wish to thank the large number of scientists and engineers who have worked tirelessly to design, build, launch, commission, calibrate, and characterize \textit{JWST}.  In particular, we appreciate the guidance of our instrument scientist supports, Tracy Beck, Armin Rest, and Swara Ravindranath.  NR \& AH thank Matt Hayes for providing a catalog of low-z analog properties and \lya\ measurements.  T.N acknowledge support from Australian Research Council Laureate Fellowship FL180100060. GPL and CM acknowledge support
by the VILLUM FONDEN under grant 37459.
KB is supported in part by the Australian Research Council Centre of Excellence for All Sky Astrophysics in 3 Dimensions (ASTRO 3D), through project number CE170100013. MB acknowledges support from the Slovenian national research agency ARRS through grant N1-0238.

\end{acknowledgements}


\bibliography{ref}
\bibliographystyle{aasjournal}
\end{document}